\documentclass[twocolumn,showpacs,preprintnumbers,amsmath,amssymb]{revtex4}
\usepackage{tabularx,graphicx}
\usepackage{color}
\begin{document}
%\documentstyle[aps]{revtex}
%\documentstyle[preprint,aps]{revtex}
%\begin{document}
\newcommand{\beq}{\begin{equation}}
\newcommand{\eeq}{\end{equation}}
\newcommand{\beqn}{\begin{eqnarray}}
\newcommand{\eeqn}{\end{eqnarray}}
\newcommand{\bmath}{\begin{subequations}}
\newcommand{\emath}{\end{subequations}}
\newcommand{\rb}{\bar{r}}
\newcommand{\bk}{\bold{k}}
\newcommand{\bkp}{\bold{k'}}
\newcommand{\bq}{\bold{q}}
\newcommand{\bkb}{\bold{\bar{k}}}
\newcommand{\br}{\bold{r}}
\newcommand{\brp}{\bold{r'}}
\newcommand{\vp}{\varphi}
\newcommand{\red}{\textcolor{red}}
%\newcommand{\red}{\textcolor{black}}

%\draft
\title{Superconductivity in the elements, alloys and simple compounds}
\author{G. W. Webb$^{a}$, F. Marsiglio$^{b}$,  J. E. Hirsch$^{a}$\footnote{Tel.: +1 858 534 3931, $email$: jhirsch@ucsd.edu}  }
\address{$^{a}$Department of Physics, University of California, San Diego, 
La Jolla, CA 92093-0319\\$^{b}$Department of Physics, University of Alberta, Edmonton,
Alberta, Canada T6G 2J}

\begin{abstract} 
We give a brief  review of superconductivity at ambient pressure in elements, alloys, and simple three-dimensional compounds.
Historically these were the first superconducting materials studied, and based on the experimental knowledge gained from them  the BCS theory of superconductivity was developed in 1957. Extended to include the effect of phonon retardation, the theory is believed
to describe the subset of superconducting materials known as `conventional superconductors', where superconductivity is caused
by the electron-phonon interaction. These  include the elements, alloys
and simple compounds discussed in this article and several other classes of materials discussed in other articles in this Special Issue.\end{abstract}
\maketitle 

\section{introduction}

%\subsection{Historical Developments in Materials}

Superconductivity was discovered by Kamerlingh Onnes  in   1911 in $Hg$ \cite{hg}, and  in $Pb$ and $Sn$ within the next two years \cite{pbsn}. 
By 1932, $Tl$, $In$, $Ga$, $Ta$, $Ti$, $Th$ and $Nb$ had also been found to be superconductors \cite{meissner}. By 1935, 15 superconducting elements were known \cite{smith},  19 by 1946 \cite{justi}, 22 by 1954 \cite{eisenstein}. Today, 31 elements are known to be superconducting at ambient pressure \cite{buzea,liexp}, 
many more at high pressures \cite{hamlinshimizu}.
Critical temperatures of the elements at ambient pressure range from $0.0003$ K for $Rh$ to $9.25$ K for $Nb$.

\begin{table}
\caption{Some superconducting alloys and compounds known in 1935 \cite{smith}.}
\begin{tabular}{l | c | c | c | c  }
  \hline
    \hline
Material  & $T_c  $   &  Material &$T_c $     \cr
 \hline  \hline
$Bi_6Tl_3$ & $6.5 K$ & $TiN$& $1.4K$   \cr
$Sb_2Tl_7$ & $5.5$ & $TiC$& $1.1$   \cr
$Na_2Pb_5$ & $7.2$ & $TaC$& $9.2$   \cr
$Hg_5Tl_7$ & $3.8$ & $NbC$& $10.1$   \cr
$Au_2Bi$ & $1.84$ & $ZrB$& $2.82$   \cr
$CuS$ & $1.6$ & $TaSi$& $4.2$   \cr
$VN$ & $1.3$ & $PbS$& $4.1$   \cr
$WC$ & $2.8$ & $Pb-As$ alloy& $8.4$   \cr
$W_2C$ & $2.05$ & $Pb-Sn-Bi$& $8.5$   \cr
$MoC$ & $7.7$ & $Pb-As-Bi$& $9.0$   \cr
$Mo_2C$ & $2.4$ & $Pb-Bi-Sb$& $8.9$   \cr
 \hline
 \end{tabular}
\end{table}

Shortly after superconductivity in $Hg$ was discovered in 1911, alloys of $Hg Au$, $Hg Cd$  $Hg Sn$ and $Pb Sn$ were also measured and found to be superconducting \cite{pbsn}.  By 1932 \cite{meissner}, a large number of binary alloys and compounds had been found to be superconducting   including $Au_2Bi$, with both elements non-superconducting \cite{aubi}. It
was also found that  when alloying a non-superconducting metal with a superconducting one $T_c$ may be increased. 
Superconducting binary compounds with one of the elements nonmetallic were found \cite{meissner}, e.g. $NbC$, with $T_c=10.1$ K, a non-superconducting metal with an insulator, $CuS$, $T_c=1.6$ K \cite{cus} and many other binary compounds, particularly
sulfides, nitrides and carbides \cite{meissner}. These early findings demonstrated   that superconductivity is a property of the solid, not of the elements forming the solid. 
Table 1 gives examples of superconducting compounds discussed in a 1935 review \cite{smith}. 

These experimental results indicated that the energy scale associated with superconductivity was of
order $k_B T_c \sim 10^{-4}$ eV.   On the other hand, it was generally believed at the time that
superconductivity originated from the electron-electron interaction neglected in Bloch's theory of electrons in single-particle energy bands. Thus a major puzzle was to understand
how an interaction many orders of magnitude larger could give rise to the low $T_c$'s measured experimentally.

\begin{table}
\caption{Critical temperature and Debye temperature of superconducting elements known in 1946 \cite{justi}.}
\begin{tabular}{l | c | c | c  }
  \hline
    \hline
Metal  & $T_c  $   &  $\theta_D$      \cr
 \hline  \hline
$Nb$ & $9.22$ & $184$   \cr
$Pb$ & $7.26$ & $86$   \cr
$La$ & $4.71$ & $?$   \cr
$Ta$ & $4.38$ & $246$   \cr
$V$ & $4.3$ & $69$   \cr
$Hg$ & $4.12$ & $69$   \cr
$Sn$ & $3.69$ & $180$   \cr
$In$ & $3.37$ & $150$   \cr
$Tl$ & $2.38$ & $100$   \cr
$Ti$ & $1.81$ & $400$   \cr
$Th$ & $1.32$ & $200$   \cr
$U$ & $1.25$ & $141$   \cr
$Al$ & $1.14$ & $305$   \cr
$Ga$ & $1.07$ & $125$   \cr
$Re$ & $0.95$ & $283$   \cr
$Zn$ & $0.79$ & $230$   \cr
$Zr$ & $0.70$ & $288$   \cr
$Cd$ & $0.54$ & $158$   \cr
$Hf$ & $0.35$ & $?$   \cr
 \hline
 \end{tabular}
\end{table}

In Table II we list the 19 superconducting elements known by the year 1946, from a paper by E. Justi \cite{justi}. The table also gives the Debye temperatures as given in that paper. It is interesting
that Justi discusses in this paper the possible effect of the ionic mass and Debye temperature on the critical temperature. He reasoned that because lattice vibrations give rise to Ohmic resistance, one might 
expect a connection between Debye temperature and superconducting $T_c$. However, from the data in Table II he concluded that there is no relation between $\theta_D$ and $T_c$ \cite{justi}. 
In addition he discussed an experiment   performed in 1941 \cite{justi2} attempting to detect any difference in the critical temperature of the two $Pb$ isotopes
$^{206}Pb$ and $^{208}Pb$ and finding identical results to an accuracy $1/1000$. From these observations he concluded in 1946  that the ionic  mass has no influence on superconductivity.

The possible relation between Debye temperature and superconducting critical temperature was also examined by  de Launay and Dolecek in 1947 \cite{isotope2}. In their paper ``Superconductivity and the Debye characteristic temperature'' they plotted the critical temperature versus Debye temperature. From this they concluded that electronegative elements have $T_c$'s well above the
$T_c$'s of electropositive elements of comparable Debye temperatures, except in the range of lowest Debye temperatures where they converge. Combining these data with 
the atomic volumes they predicted that, at atmospheric pressure, Scandium and Yttrium  should not be superconducting (correct) and that Ce, Pr and Nd  should be superconducting (incorrect).

\begin{table}
\caption{Critical temperature, Debye temperature, atomic mass, measured and calculated isotope exponents of superconducting elements.
Measured values are taken from a table in Ref. [\onlinecite{isotope}] and theoretical values are taken from a table in Ref. [\onlinecite{garland63}].}
\begin{tabular}{l | c | c | c |c | c| c }
  \hline
    \hline
Metal  & $T_c  $   &  $\theta_D$  &M&  $\alpha$ &$\alpha_{theory}$  \cr
 \hline  \hline
$Nb$ & $9.25$ & $275$  & 93 &  & \cr
$Tc$ & $8.2$ & $450$   &99  &  & \cr
$Pb$ & $7.2$ & $105$ &207  & 0.48 & 0.47 \cr
$La$ & $6$ & $142$   & 139 &  & \cr
$V$ & $5.4$ & $380$  & 51 &  & 0.15 \cr
$Ta$ & $4.4$ & $240$  &181  &  & 0.35 \cr
$Hg$ & $4.15$ & $72$   & 201 &0.5  & 0.465 \cr
$Sn$ & $3.7$ & $200$  & 119 & 0.46 & 0.44\cr
$In$ & $3.4$ & $108$  & 115 &  & \cr
$Tl$ & $2.4$ & $78.5$  & 204 & 0.5 & 0.445 \cr
$Re$ & $1.7$ & $430$  &  186&  0.38& 0.3\cr
$Th$ & $1.4$ & $163$  & 232 &  & \cr
$Pa$ & $1.4$ & $?$   & 231 &   &\cr
$U$ & $1.3$ & $207$  & 238 & -2 & \cr
$Al$ & $1.18$ & $428$  & 27 &  & 0.345 \cr
$Ga$ & $1.08$ & $320$ &70  &  & \cr
$Am$ & $1$ & $154$  & 243 &   &\cr
$Mo$ & $0.92$ & $450$  & 96 & 0.37 & 0.35 \cr
$Zn$ & $0.85$ & $327$   & 65 & 0.3 & \cr
$Os$ & $0.7$ & $500$  & 190 & 0.21 & 0.1\cr
$Zr$ & $0.6$ & $291$  & 91 & 0 & 0.35 \cr
$Cd$ & $0.52$ & $209$  & 112 & 0.5 & 0.365\cr
$Ru$ & $0.5$ & $600$   &  101& 0 & 0.0\cr
$Ti$ & $0.5$ & $420$  & 48 &   & 0.2\cr
$Hf$ & $0.38$ & $252$  & 176 &  & 0.3\cr
$Ir$ & $0.1$ & $420$   & 192 &  & -0.2 \cr
$Lu$ & $0.1$ & $210$   &139  &  & \cr
$Be$ & $1440$ & $1440$ &9  &  & \cr
$W$ & $0.01$ & $400$  &184  & &  \cr
$Li$ & $0.0004$ & $344$ & 7 & &  \cr
$Rh$ & $0.0003$ & $480$ &103  &  & \cr
 \hline
 \end{tabular}
\end{table}

In view of these investigations it is remarkable that just three years later in 1950 Herbert Fr\"ohlich proposed \cite{frohlich}  that  superconducting critical temperatures
should be proportional to $M^{-\alpha}$, with $M$ the ionic mass and $\alpha=0.5$ the isotope exponent. 
This was done without knowledge \cite{view1,view2,view3} of the isotope effect experiments \cite{maxwell,serin} being
conducted at the same time   that measured an isotope exponent $\alpha\sim 0.5$ in $Hg$ and shortly thereafter in $Pb$ \cite{pb},  $Sn$ and $Tl$ \cite{sn}. 
Table III lists the isotope exponents of these and several other elements measured since then \cite{isotope,garland63}.

After the experimental findings of an isotope effect, the focus of theoretical efforts to understand the origin of the interaction leading to superconductivity shifted from the
electron-electron interaction to the electron phonon interaction. In 1957   BCS developed their theory based on an effective instantaneous attractive interaction between electrons 
mediated by phonons \cite{bcs}, that also predicts $\alpha=0.5$. BCS theory, extended to take into account the fact that the effective interaction between electrons mediated by phonons is not
instantaneous but retarded, is believed to describe the superconductivity of all elements at ambient pressure, and of thousands of superconducting compounds.
The tabulation by Roberts (1976) \cite{roberts} lists
several tens of thousands of superconducting alloys and compounds, almost all with critical temperatures below $20K$, believed to be described by BCS theory. 

\section{Response to a Magnetic Field: Phenomenology}

%fig. 0
 \begin{figure}
%\resizebox{6.5cm}{!}{\includegraphics[width=7.5cm]{hc.pdf}}
%\resizebox{6.1cm}{!}{\includegraphics[width=7.5cm]{fig03_hc2.pdf}}
\resizebox{6.5cm}{!}{\includegraphics[width=7.5cm]{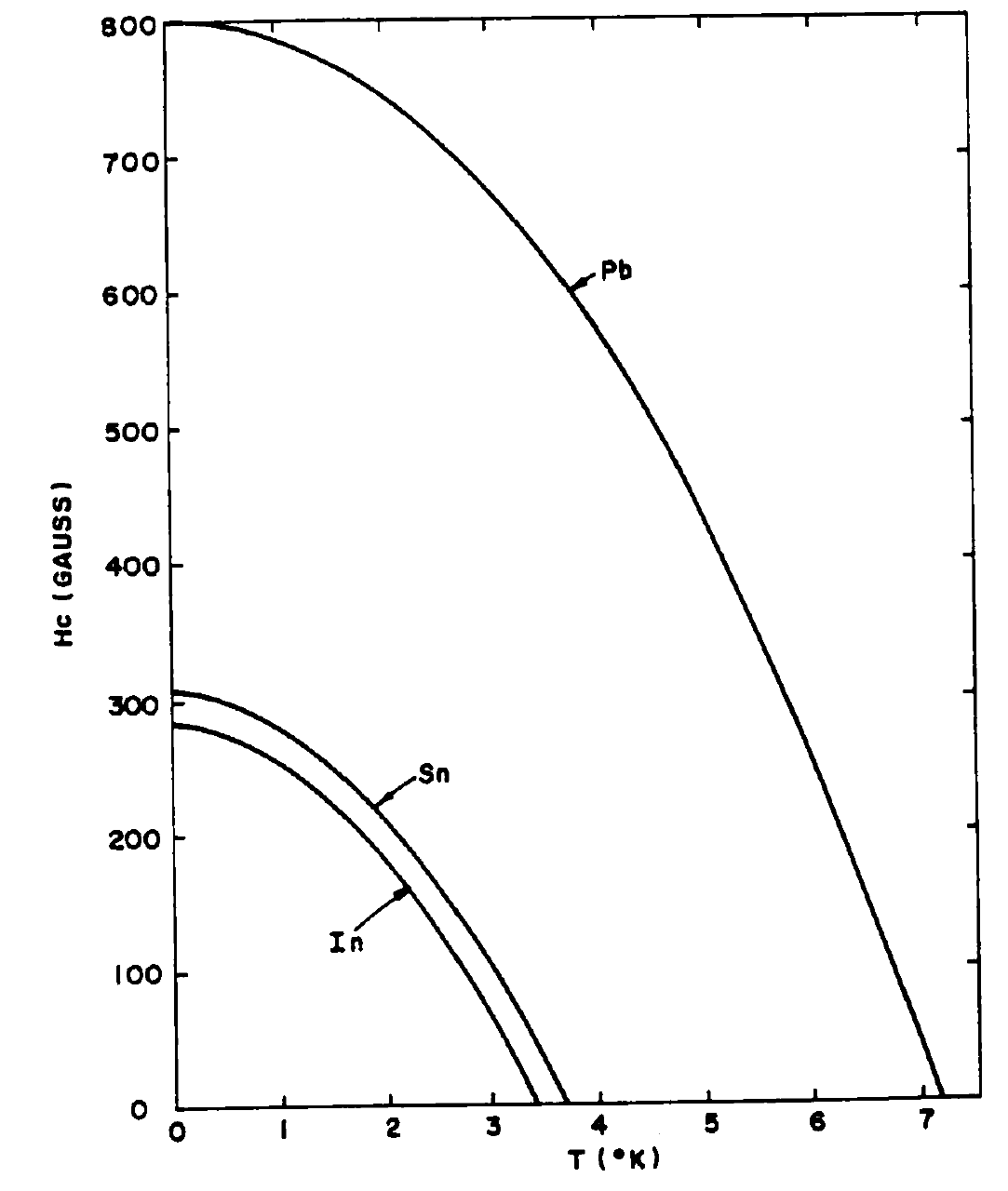}}
\resizebox{6.1cm}{!}{\includegraphics[width=7.5cm]{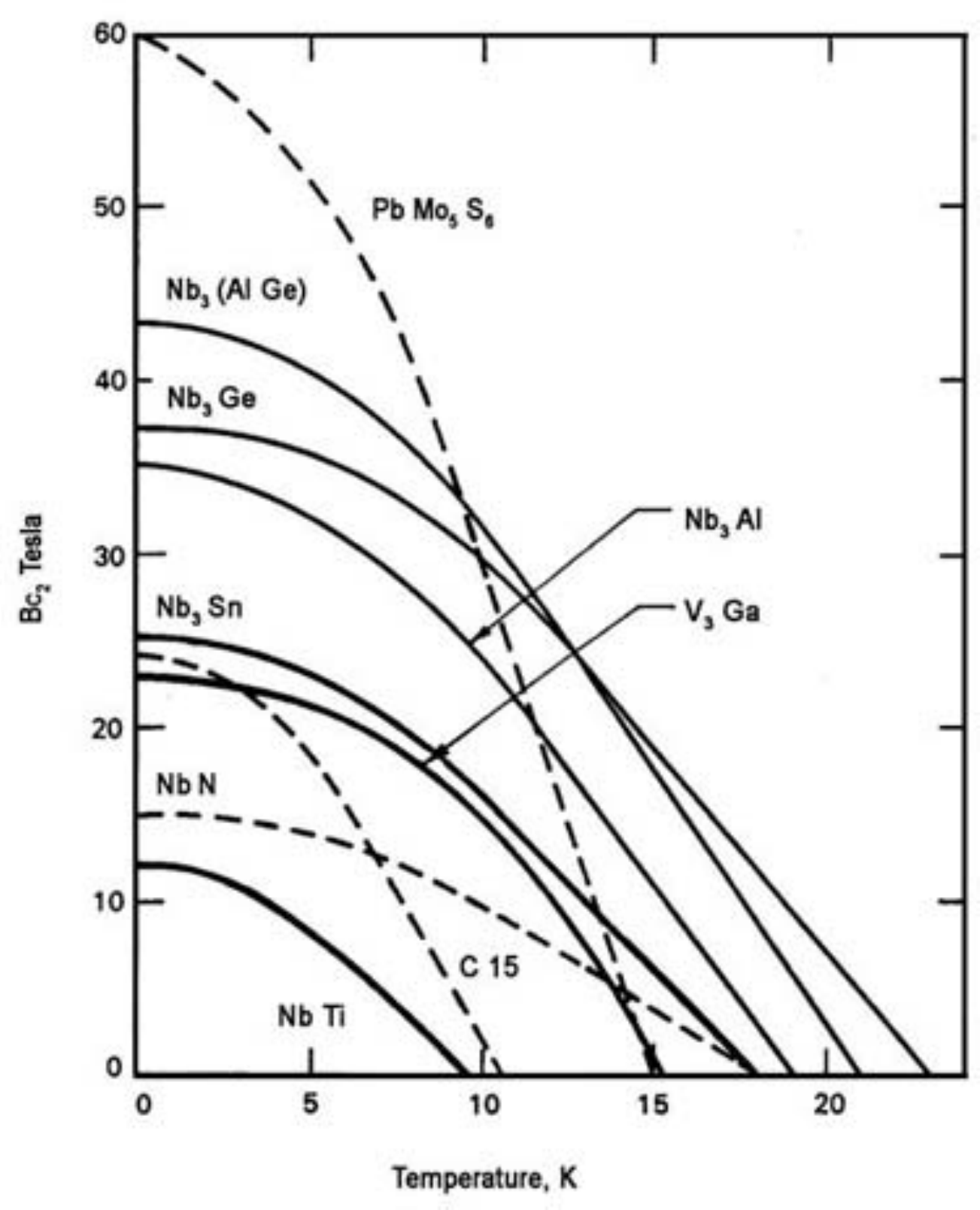}}
\caption {(a) Upper panel shows the $H_c$ vs temperature curves for three elemental type-I superconductors. (b) Lower panel
shows the upper critical field, $H_{c2}$ for several superconducting compounds, belonging to various families. Note the huge
difference in critical magnetic field strengths between type-I (a) and type-II (b) superconductors. Figure in (a) adapted from Ref.~[\onlinecite{cody73}]
and figure in (b) used from Tables of Physical \& Chemical Constants (16th edition 1995). 2.1.4 Hygrometry. Kaye \& Laby Online. Version 1.0 (2005).}
\label{figure0}
\end{figure} 

Much of the focus for superconducting materials is on increasing $T_c$. This is of course important for applications, but as Geballe {\it et al.} \cite{geballe}
emphasize, it is also a primary measure of our understanding of the mechanism for superconductivity. In contrast the response of a material to an 
applied magnetic field is more generic, in the sense that a microscopic theory is usually not required to understand this response. In fact the Ginzburg-Landau 
theory \cite{ginzburg50} often suffices to provide a detailed description of the magnetic state, whether the material is type-I or type-II. In a type-I superconductor
the magnetic response is perfect diamagnetism, with the magnetic field completely expelled provided the field strength is less than a critical value, $H_c$. At this field
value the material reverts to the normal state. In a type-II superconductor, the material exhibits perfect diamagnetism
up to a critical field $H_{c1}$; with increasing applied field, flux begins to penetrate the material in the form of vortices. This continues to occur up to an
upper critical field, $H_{c2}$, after which they become
normal \cite{remark0}.

Since $H_{c2} >> H_{c}$ (by several orders of magnitude), type-II superconductors are most useful in applications that have a magnetic field
present. Whether a material is a type-I or a type-II superconductor depends on the so-called Ginzburg-Landau parameter, $\kappa \equiv \lambda/\xi$,
where $\lambda$ is the penetration depth and $\xi$ is the superconducting coherence length. Since the coherence length can decrease with a decreased 
scattering length, then a type-I superconductor can be made into a type-II superconductor through disorder. There are approximately 30 pure elements that
superconduct at atmospheric pressure; three of these, $Nb$, $V$, and $Tc$ are type-II while the rest are type-I. Essentially all compounds are type-II.
In Fig.~\ref{figure0} we show some experimental data for the critical fields of (a) a few type-I elemental superconductors, and (b) a few type-II superconducting
compounds. Note that while the temperature scale in (b) is about a factor of 3 higher than in (a), the magnetic field strengths in (b) about 1000 times
higher than in (a).

\section{BCS theory and its extensions (Eliashberg)}

The papers in this Special Issue each deal with a particular family of superconductor. By design they focus on the materials and experimental
properties, with limited theoretical discussion. As Bernd Matthias said it in the famous `Science' debate with Philip Anderson \cite{anderson64},
we wanted to focus on `The Facts'. Nonetheless, as the reader will see from the various contributions in this Issue, it is difficult to examine
material properties without an underlying theoretical framework. For example, the McMillan equation \cite{mcmillan68} comes up in a number
of places as a means to understand trends in superconducting $T_c$. We therefore felt it would be useful to provide here a sketch of the
`conventional' theory of superconductivity.

The zero temperature BCS theory \cite{bcs} consists of a variational wave function, motivated by a collection of Cooper pairs \cite{cooper56}.
Using this wave function, and a mean field simplification at finite temperature, one arrives at the simplest form for the superconducting transition temperature, given by
\beq 
T_c =1.13 \theta _D e^{-1/[g(\epsilon_F)V]}
\label{bcs_tc}
\eeq
where $g(\epsilon_F)$ is the density of states at the Fermi energy and $V$ is the effective electron-electron attraction within a range $\hbar\omega_D
\equiv k_B \theta_D$ of
the Fermi energy. One should take special note that BCS theory is a pairing theory, and in principle, has nothing to say about pairing mechanism.
Here, following BCS \cite{bcs}, a phonon mechanism is implied by the use of a cut off energy, $k_B \theta_D$.
Many extensions of BCS theory are possible beyond this simple model, spanning minor considerations like a non-constant
density of states near the Fermi level, to more serious modifications like inhomogeneities (leading to the Bogoliubov-de Gennes (BdG) equations
\cite{degennes66}), or an order parameter with nodes, or significant retardation effects (leading to Eliashberg theory \cite{eliashberg60}). 
In discussing superconductivity amongst the elements, Eliashberg theory is required for a quantitative understanding of many of the
superconducting properties, so we will expand in this direction below.

BCS theory alone allows us to understand a number of simple but important properties, which we now discuss before moving on to
Eliashberg theory. First, as already mentioned, superconducting $T_c$ will have an isotope effect, and since $T_c \propto \theta_D$,
then $T_c \propto M^{-\alpha}$ with $\alpha = 0.5$. As mentioned already by Geballe et al. \cite{geballe}, even in the absence of
theoretical motivation, Kamerlingh Onnes and Tuyn \cite{onnes23} looked (unsuccessfully) for an isotope effect in Pb in 1923, 
as did Justi \cite{justi2} 18 years later; then one was
found in 1950 in Hg \cite{maxwell,serin}. BCS theory predicts an energy gap in the single particle density of
states; this was confirmed by tunneling measurements a number of years later \cite{giaever60}. Finally, one of the non-intuitive confirmations
of BCS theory is the observation of the so-called Hebel-Slichter coherence peak in the NMR relaxation rate of
Aluminum \cite{hebel57,redfield59}, where the relaxation rate rises initially as the temperature is lowered below $T_c$, before becoming
suppressed due to the opening of a gap.

%Applications of superconductivity, such as those that utilize SQUIDS, are not the subject of this article or any others in this Special Issue,
%but in the present context, they constitute indirect evidence for the validity of the BCS framework, since the Josephson Effect \cite{josephson62}
%is a derivative of this framework. Applications such as these, however, have little to say concerning the mechanism for superconductivity
%and/or the search for higher $T_c$ materials.

%fig. 2
 \begin{figure}
%\resizebox{7.5cm}{!}{\includegraphics[width=7.5cm]{black_march_al(a).pdf}}
%\resizebox{7.5cm}{!}{\includegraphics[width=7.5cm]{black_march_al(b).pdf}}
\resizebox{7.5cm}{!}{\includegraphics[width=7.5cm]{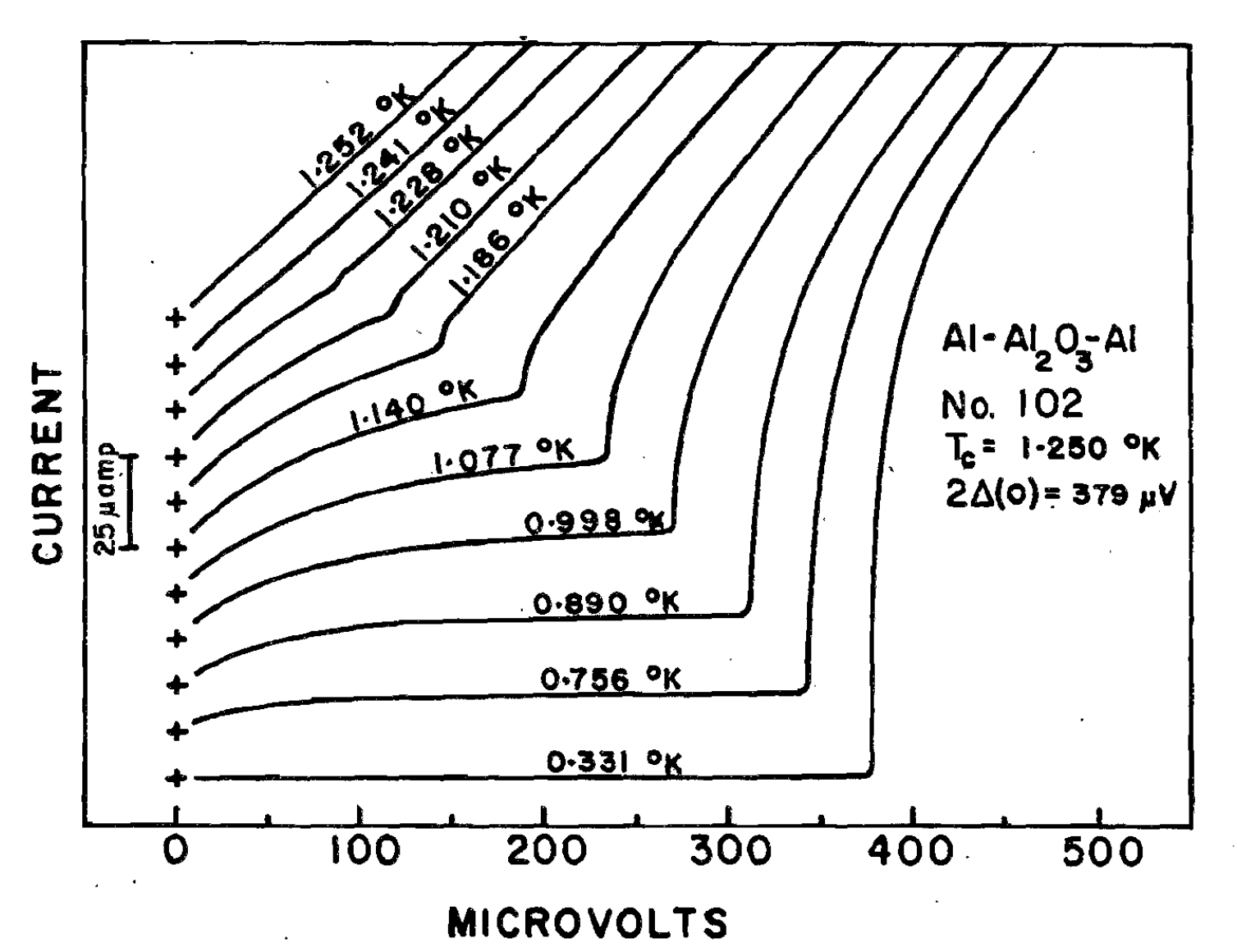}}
\resizebox{7.5cm}{!}{\includegraphics[width=7.5cm]{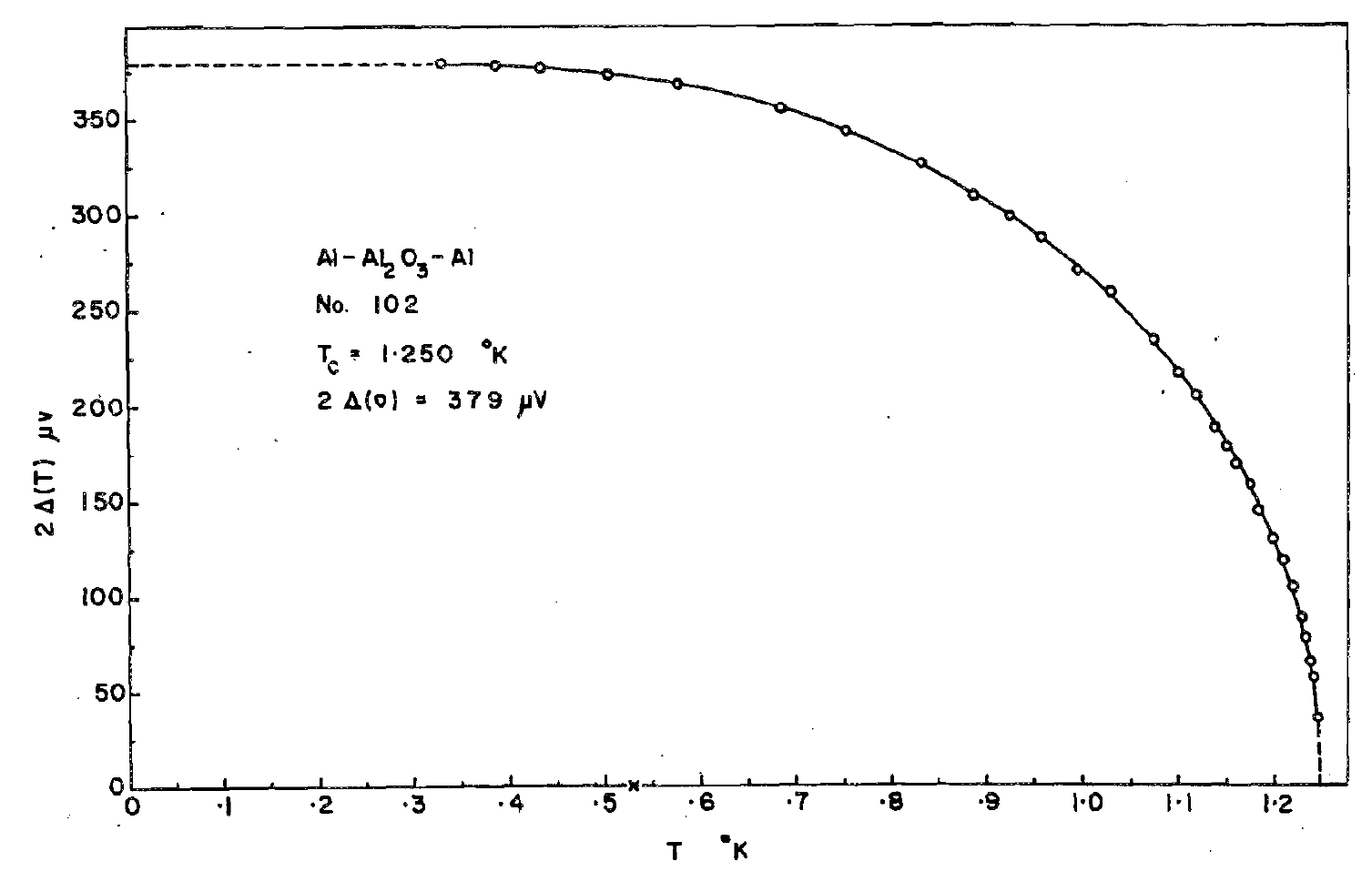}}
\caption {(a) IV characteristics for Al-I-Al junctions, and (b) the resulting normalized
superconducting gap as a function of reduced temperature (points) compared with
BCS theory (curve). The agreement is very good. From Ref.~[\onlinecite{black68}]. }
\label{figure1}
\end{figure} 
%fig. 3
 \begin{figure}
%\resizebox{7.5cm}{!}{\includegraphics[width=7.5cm]{spec_heat_al_bcs.pdf}}
\resizebox{7.5cm}{!}{\includegraphics[width=7.5cm]{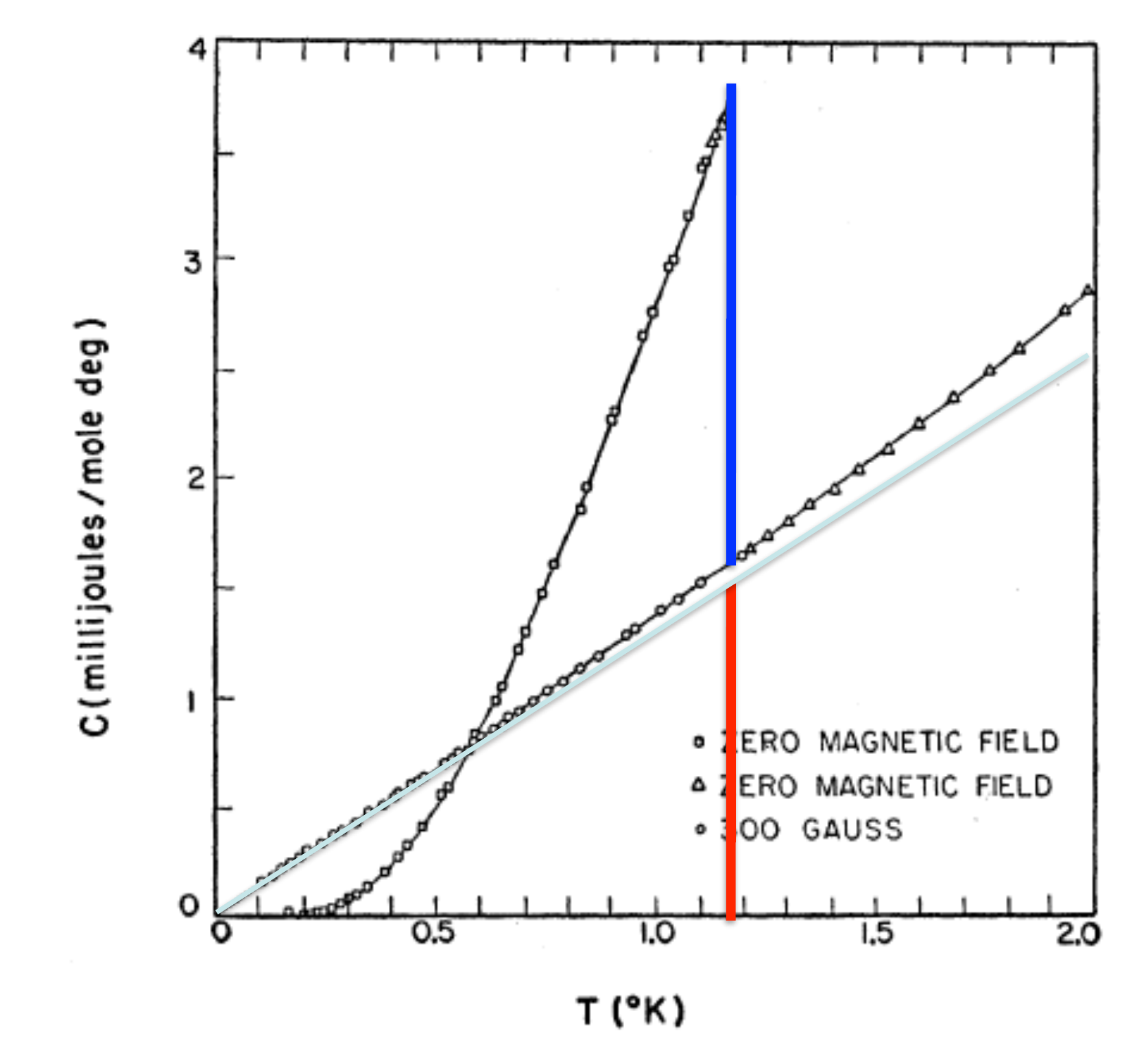}}
\caption {Specific heat measurements in both the normal and superconducting states
for Al (from Ref.~[\onlinecite{phillips59}]). Normal state results are achieved by the application of a magnetic field of 300 Gauss.
The lightly shaded line and the two vertical lines have been added to indicate
the normal state electronic specific heat ($\gamma T$) (grey) and the normal state electronic contribution at $T_c$ ($\gamma T_c$) (red) 
and additional jump at $T_c$ as predicted by BCS theory ($\Delta C_{es} = 1.43 \gamma T_c$) (blue),
respectively. The data in the superconducting state is in very good agreement (slightly lower) than the BCS 
weak coupling prediction. Adapted from Ref.~[\onlinecite{phillips59}]. }
\label{figure2}
\end{figure} 

Examples of some experiments with excellent agreement with BCS theory are the tunneling measurements for Al-I-Al junctions 
(see Fig.~\ref{figure1}) and specific heat measurements on Al (see Fig.~\ref{figure2}). There are many others in the literature \cite{parks69}.
It is clear from these examples that Aluminum is the `poster child' for BCS weak coupling theory. Nonetheless, even amongst the elemental
superconductors there exist so-called `bad actors' whose properties clearly do not conform quantitatively to BCS theory. Eliashberg
theory was explored in part because of these discrepancies, and the `poster child' for Eliashberg theory is Lead. Many reviews 
\cite{scalapino69,mcmillan69,allen82,rainer86,carbotte90,marsiglio08}, have been written on this subject, so here we will highlight
some of the experimental manifestations. Note that Eliashberg theory is sometimes called the strong coupling version of BCS theory; this
is somewhat of a misnomer, as both are developments with Fermi Liquid Theory as a starting point, and the term `strong coupling' is
generally reserved for situations in which kinetic energy (and therefore Fermi Liquid ideas) is initially ignored. It is more accurate to
refer to Eliashberg theory as an extension of BCS theory with retardation effects properly taken into account \cite{remark1}.

The order parameter in Eliashberg theory becomes frequency dependent and complex. Both of these complications result from
retardation effects. One of the immediate manifestations of this theory is a series of non-universal results for various properties
that are universal within BCS theory. But even the theory for $T_c$ becomes more complicated, as epitomized, for example, by
the McMillan equation \cite{mcmillan68,allen75} for $T_c$:
\beq
T_c = {\hbar \omega_{\rm \ell n} \over 1.2k_B} \exp{\biggl({-1.04(1+\lambda) \over \lambda - \mu^\ast (1 + 0.62 \lambda)}\biggr)}
\label{mcmillan}
\eeq
where $\omega_{\rm \ell n}$ is used as an average phonon frequency, and it and $\lambda$ are defined by
\beq
\omega_{\ln} \equiv \exp{\left[ {2 \over \lambda}
\int_0^\infty d\nu \, \ln{(\nu)} {\alpha^2(\nu)F(\nu) \over \nu}  \right]}
\label{wlog}
\eeq
and
\beq
\lambda \equiv 2
\int_0^\infty d\nu \, {\alpha^2(\nu)F(\nu) \over \nu}.
\label{lambda}
\eeq
Both of these parameters are related to moments of the so-called Eliashberg function, $\alpha^2(\nu)F(\nu)$; this
function describes the modes of excitations (in this case phonons) through which electrons effectively
attract one another. 
They do this by emitting virtual phonons, in analogy to the photon exchange for the ordinary Coulomb interaction.
But phonon propagation is several orders of magnitude slower than photon propagation, so properly accounting for
this time delay means one electron attracts the other not to itself, but to where it used to be.
%, as depicted in the cartoon shown in Fig.~\ref{figure3}. See the figure caption for a simple
%explanation. 
%%fig. 3
% \begin{figure}
%\resizebox{6.0cm}{!}{\includegraphics[width=7cm]{fig3a.pdf}}
%\resizebox{6.0cm}{!}{\includegraphics[width=7cm]{fig3b.pdf}}
%\caption {(a) An electron (shown in red in the central region of four neighbouring positively charged ions) polarizes
%the surrounding ions (excites a phonon) to create a slightly more positive environment. (b) A short instant later, the
%electron has left the region, which now remains positively charged, and draws in a second electron. Effectively, the
%first electron has attracted the second electron to where it used to be. This partially accounts for the fact that a BCS
%description (in momentum space) works as well, since this appears to be a long range interaction between the two
%electrons, even though all of the electron-ion interactions take place locally.}
%\label{figure3}
%\end{figure} 
%The `dynamics' depicted in Fig.~\ref{figure3} also accounts for the presence (and smallness) of $\mu^\ast$. 
This `dynamics' also accounts for the smallness of the direct Coulomb interaction between two electrons, depicted by $\mu^\ast$. 
%This parameter represents the direct Coulomb repulsion experienced between the two electrons. 
This repulsion would be
overwhelmingly large, except that the two electrons are not in the same place at the same time, when they best take advantage of the 
virtual phonon exchange. This
diminishing effect of the direct Coulomb potential is crucial for phonon-mediated superconductivity, and is known as the 
pseudo potential effect \cite{bogoliubov58,morel62}, with an expression given by
\beq
\mu^\ast = {\mu \over 1 + \mu \ln{({\epsilon_F \over \hbar \omega_D})}},
\label{mustar}
\eeq
with $\epsilon_F$ the Fermi energy and $\mu = g(\epsilon_F) U$ the dimensionless `bare' Coulomb interaction. Typically
$\epsilon_F >> \hbar \omega_D$, and so $\mu^\ast << \mu$, with a limiting value of $1/\ln{(\epsilon_F/(\hbar \omega_D)}$. 
This scaling of the Coulomb repulsion is also responsible for making
calculations more tractable, as frequencies out to several (say, 6) times the phonon energy scale are required (about 60 meV for Lead), 
compared with several times the electronic bandwidth (about 2 orders of magnitude higher). A simple model illustrating this
can be found in Ref. [\onlinecite{marsiglio92}].

%fig. 4
 \begin{figure}
%\resizebox{7.5cm}{!}{\includegraphics[width=7cm]{gap_ratio_vs_tcwln_mitrovic.pdf}}
\resizebox{7.5cm}{!}{\includegraphics[width=7cm]{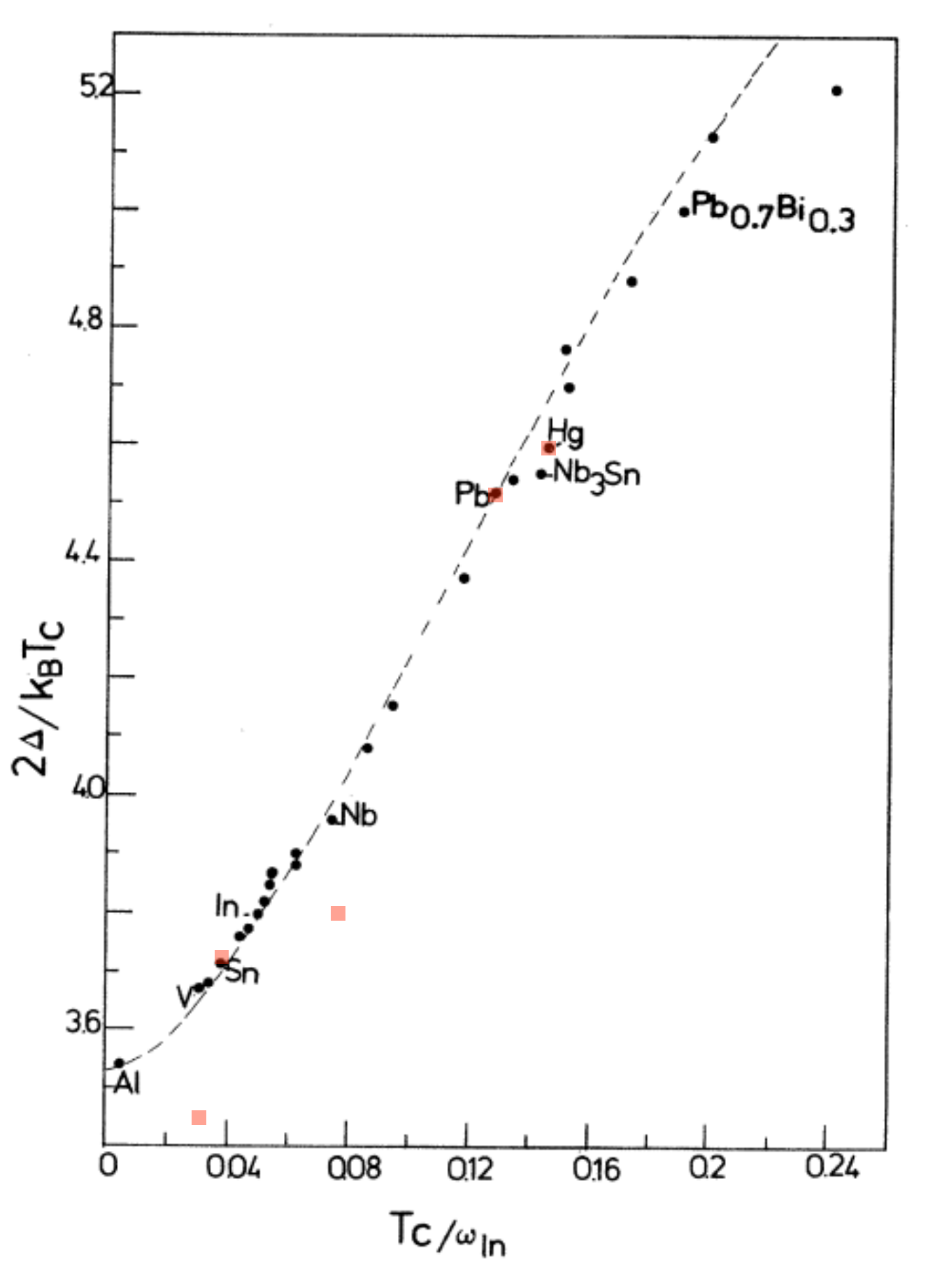}}
\caption {The gap ratio $2\Delta_0/(k_BT_c)$ as a function of $T_c/\omega_{\rm \ell n}$. The black circles indicate theoretical
calculations, with some of the elements and a couple of binary alloys indicated. The unmarked circles refer
mostly to various binary alloys \cite{mitrovic84}. These calculations use an electron-phonon spectral function 
$\alpha(\nu)^2F(\nu)$ and value of $\mu^\ast$ extracted from tunneling experiments, or, in some cases taken from calculations 
\cite{carbotte67,carbotte68}. Selected experimental values  are indicated with red squares. Note the excellent agreement of theory
with experiment in the case of Sn, Pb and Hg, with more deviation in the case of Vanadium and Niobioum. Sources are
available in Ref.~[\onlinecite{mitrovic84}]. Figure is taken and then adapted from Ref.~[\onlinecite{mitrovic84}].}
\label{figure3}
\end{figure} 

Damping effects are essentially left out of simplifications like the McMillan equation, except for the presence of the mass enhancement
factor, $1 + \lambda$, in the numerator of the exponential. This tells us that the electron does become heavier as a result of the
electron-phonon interaction, and $m^\ast/m \approx 1 + \lambda$ is essentially the weak coupling remnant of the polaronic 
mass enhancement.

Full solutions of the Eliashberg equations display non-universality of various dimensionless quantities as a function of retardation effects.
Mitrovi\'c et al. \cite{mitrovic84} identified a dimensionless parameter that grows from zero with increasing retardation effects; this is
$T_c/\omega_{\rm \ell n}$. As this parameter tends to zero, various superconducting properties tend to their BCS limit. An example is
the gap ratio, $2\Delta_0/(k_BT_c)$, and a plot of this property vs. $T_c/\omega_{\rm \ell n}$ is shown in Fig.~\ref{figure3}, along with
some experimental data. Mitrovi\'c et al. derived an approximate expression,
\beq
{2\Delta_0 \over (k_B T_c)} = 3.53 \biggl[ 1 + 12.5 \bigl({T_c
\over \omega_{\ln} }\bigr)^2 \ln{({\omega_{\ln} \over 2 T_c})} \biggr],
\label{str_gap}
\eeq
which is also plotted as a dashed line. This simple expression clearly captures the essence of the theoretical results; note that some of 
the experimental values are in close agreement with the theoretical ones, while others remain closer to the universal BCS value.
%fig. 5
 \begin{figure}
%\resizebox{7.5cm}{!}{\includegraphics[width=5cm]{mcmillan_rowell_a.pdf}}
%\resizebox{7.5cm}{!}{\includegraphics[width=5cm]{mcmillan_rowell_b.pdf}}
%\resizebox{7.5cm}{!}{\includegraphics[width=5cm]{mcmillan_rowell_c.pdf}}
\resizebox{6.0cm}{!}{\includegraphics[width=5cm]{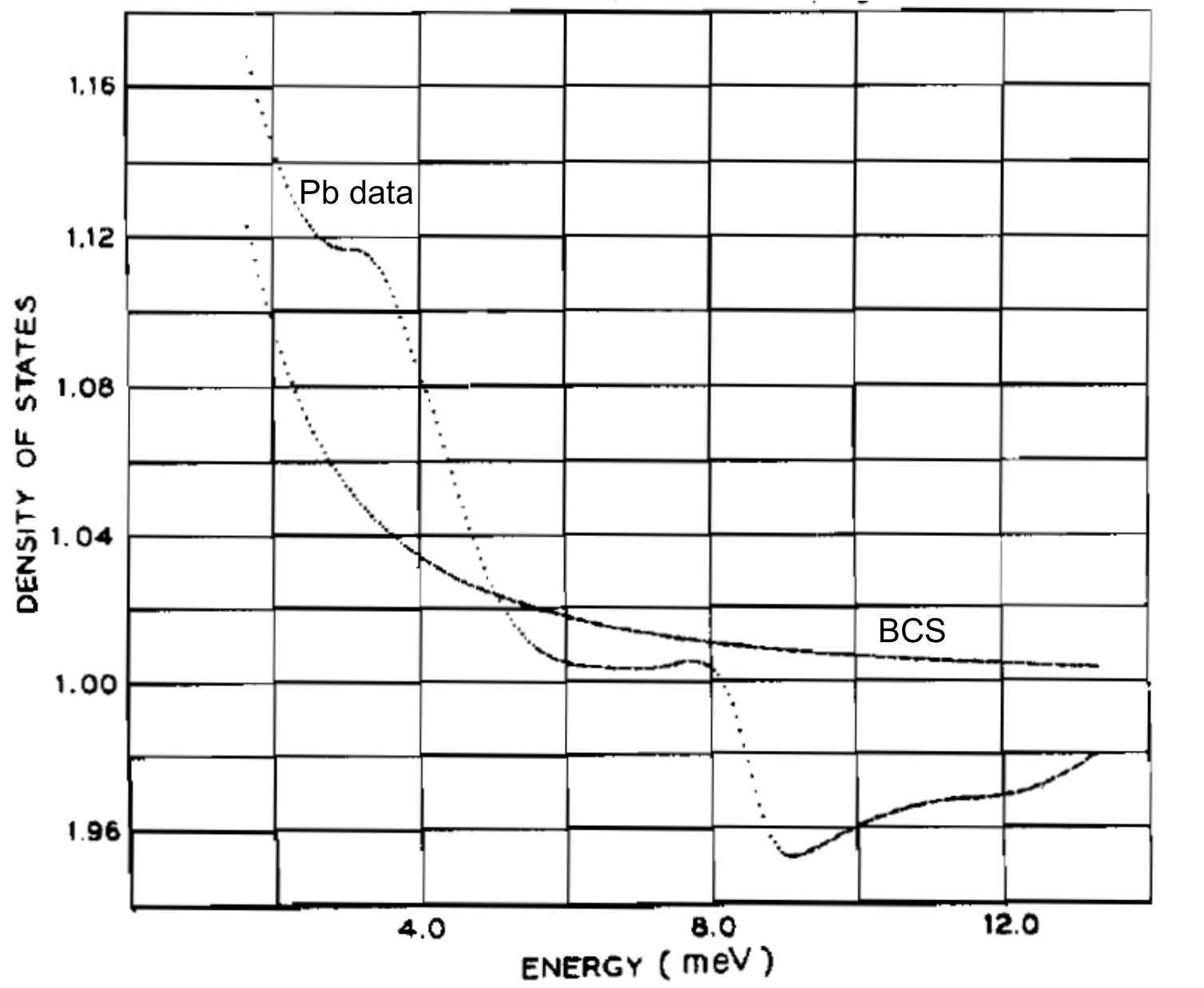}}
\resizebox{6.4cm}{!}{\includegraphics[width=5cm]{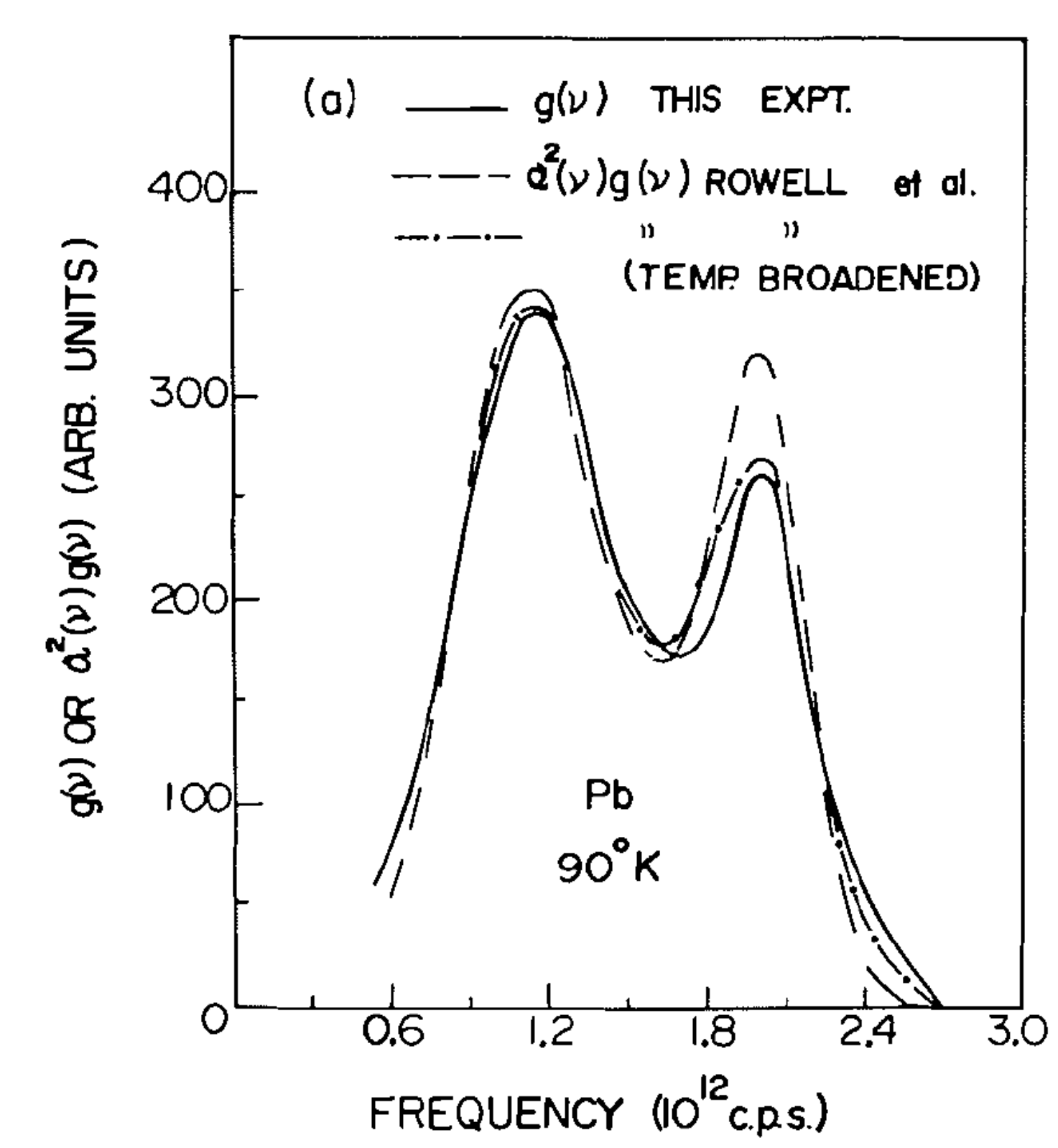}}
\resizebox{6.0cm}{!}{\includegraphics[width=5cm]{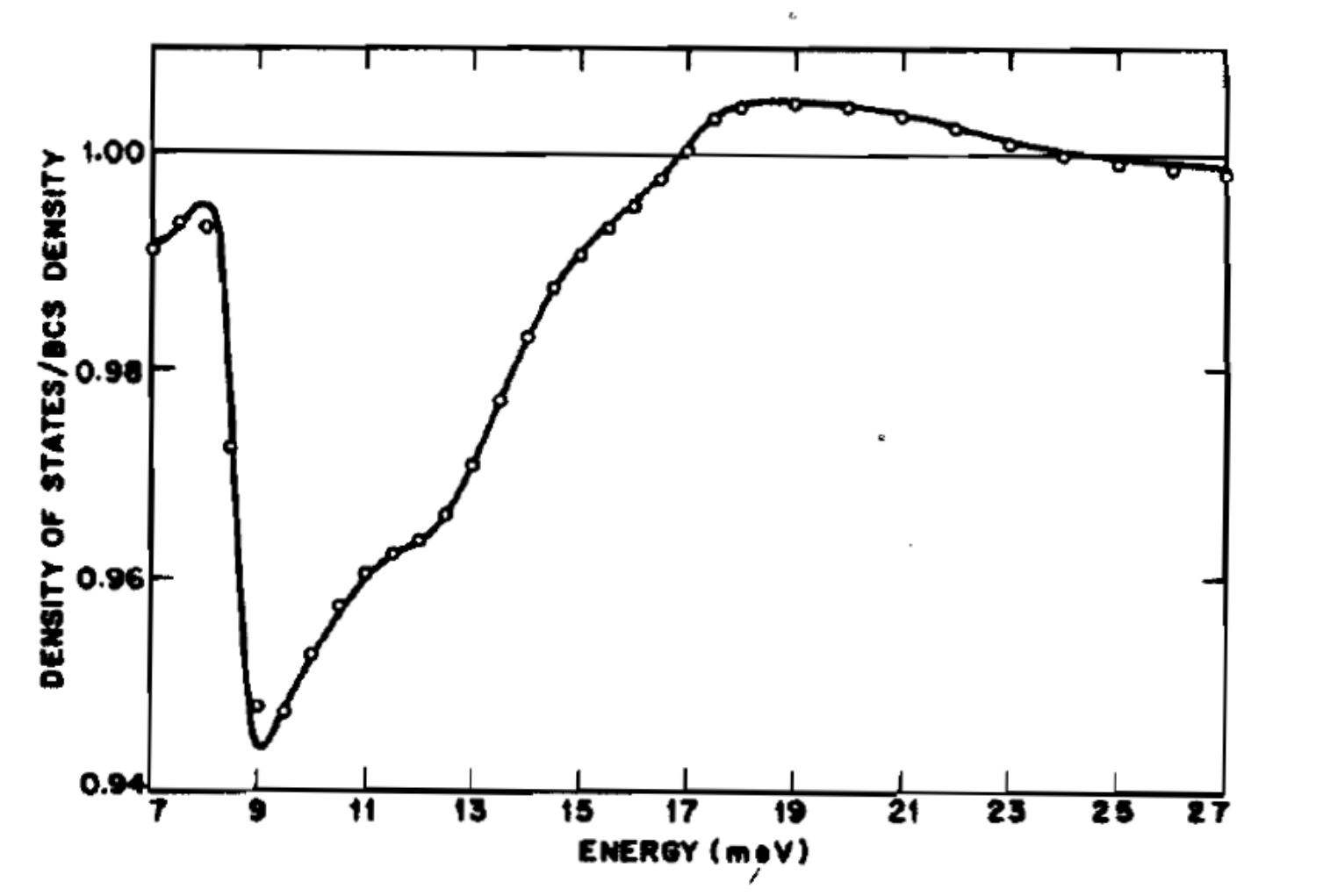}}
\caption {(a) The density of states for a Pb superconductor, obtained from conductance measurements of
a Pb-I-Pb tunnel junction \cite{mcmillan69}. The BCS theory expectation value is shown for
comparison. In (b) the extracted $\alpha^2(\nu)F(\nu)$ is shown (referred to as $\alpha^2(\nu)g(\nu)$ in the figure; this 
is obtained by demanding that the theory reproduce
exactly the observed modulations with frequency. Also superimposed is the phonon density of states (denoted $g(\nu)$
in the figure)
as measured through neutron scattering %\cite{brockhouse62}
\cite{roy70}; the rough agreement makes it clear that the excitations
responsible for the modulations are phonons. Note that a number of consistency checks all prove positive. For
example, the spectral function turns out to be positive definite (as it must), the required value of $\mu^\ast$ is positive (indicating
a competing repulsion and not an additional attractive mechanism) and finally, in part (c) a comparison of the theory (curve) and experiment (points)
in the `multiple-phonon-emission' region is shown to illustrate the predictive power of the Eliashberg theory \cite{mcmillan69}.
Figures in (a) and (c) are from Ref.~[\onlinecite{mcmillan69}] and the figure in (b) is
from Ref.~[\onlinecite{roy70}].}
\label{figure4}
\end{figure}

The strongest evidence for the applicability of Eliashberg theory to elemental superconductors comes from tunneling measurements.
Very early on observed modulations as a function of frequency in the measured current-voltage characteristics, especially in Lead, were
suspected of being due to the electron-phonon interaction. Model calculations \cite{rowell63,schrieffer63} confirmed that Eliashberg 
theory could explain these modulations, and a short while later McMillan and Rowell \cite{mcmillan65} used Eliashberg theory
to invert the tunneling data and extract $\alpha^2(\nu)F(\nu)$ and $\mu^\ast$. The latter was fit to a measurement of the tunneling gap edge,
for example. An example of the data and the spectrum extracted from this data are shown in Fig.~\ref{figure4}, and explained in that figure
caption. Further explanation is available in Ref. [\onlinecite{mcmillan69}].

These have been interpreted as being very strong indications of the validity of
Eliashberg Theory for elemental superconductors.
%These are all  indications of the validity of Eliashberg Theory for elemental superconductors. 
Probably the `Achilles heel' for
which at the very least further understanding is required is the significant reduction of the direct Coulomb repulsion, manifested in the
single number, $\mu^\ast$. 

%In the simplest electron-phonon interaction model (Holstein) the Hamiltonian has a term
%\beq
%H_{e-ph}=\alpha q_i n_i
%\eeq
%with $n_i$ the local electron density and $q_i$ a local ionic displacement, with dynamics given by
%\beq
%H_{ph}=\frac{p^2}{2M}+\frac{1}{2}Kq_i^2
%\eeq
%whence the effective electron-electron attraction  is given by $V=\alpha^2/2K$ and the Debye tempera3ture by $\theta_D=\hbar(K/M)^{1/2}/k_B$.
%
%
%more realistic models (briefly)
%
%Coulomb pseudopotential (briefly)
%
%Hopfield parameter?
%

\section{Isotope effect}

The simplest BCS prediction for the isotope effect, using Eq.~\ref{bcs_tc} is that the isotope coefficient, $\alpha = 0.5$. 
Use of Eliashberg theory does not alter this result, but in either case there will be a reduction in the isotope coefficient
due to the interplay between the electron-phonon and direct Coulomb interactions. The reason for the reduction is
simple to understand in the following way \cite{garland63}: for increased isotope mass, while the prefactor 
in Eq.~\ref{mcmillan} goes down, therefore causing a decrease in $T_c$, this is offset slightly by the fact that the overall
interaction is slightly more retarded than it was previously. This means that the electrons attract one another more effectively,
because $\omega_D$ is even lower compared to the Fermi energy than before, so that $T_c$ will increase as a result. The lower
$T_c$ is, the more effective is this mechanism, and therefore the isotope coefficient will be less than $\alpha = 0.5$ By the
time Garland performed his study in 1963, quite a number of elemental superconductors were known with very low values of
$\alpha$, most notably Ru (see Table III), and he was able to understand this very low value, along with others, based on a
competition between these two effects. A general statement is that the lower $T_c$ is, the more likely that the isotope coefficient
approaches zero. More complete calculations were performed in Ref. [\onlinecite{rainer79}] and a comparison with what is
inferred from the McMillan equation is provided in Ref. [\onlinecite{knigavko01}].

Other elemental superconductors exist where a quantitative understanding of the isotope coefficient is still lacking \cite{allen88,fowler67}.
The case of $\alpha-$uranium stands out, and has an anomalous coefficient of $\alpha=- 2$ \cite{fowler67}.

In simple compounds the situation is similar. The study in  Ref. [\onlinecite{rainer79}] was motivated by the anomalous isotope effect observed in the
Palladium-Hydride system \cite{stritzker72}, where $T_c$ {\it increases} with increasing isotope mass. The isotope effect in compounds
requires the notion of a ``differential isotope exponent'' \cite{rainer79} to determine the contribution from alterations in the electron-phonon
spectral function at different frequencies. In the case of a system where the different atoms vary considerably in mass (as in the Pd-H system)
then high frequency components can be attributed specifically to vibrations associated with the lighter mass element. Thus one can readily
determine the expected isotope effect due to only the Hydrogen-Deuterium substitution. The isotope coefficient in this case will be reduced
from $0.5$, but it will never go below zero, and thus cannot explain the experimental result \cite{rainer79}.

We should note that a theory to explain this anomaly was constructed \cite{ganguly75,klein77}, but it invoked large anharmonic effects to determine superconducting $T_c$ and the isotope coefficient $\alpha$ \cite{struzhkin15}. More recently superconductivity has been found in H$_2$S \cite{drozdov14}, 
in a system where anharmonic effects are expected to be even larger, because of the much higher temperatures involved. Here, however, the isotope coefficient does not have an anomalous sign, and is in fact {\it much higher} than expected from BCS/Eliashberg theory with harmonic phonons.

\section{Superconductivity in the elements}

It is generally believed that the 31 superconducting elements at ambient pressure listed in table III
are described by BCS-Eliashberg theory, and that the reason the remaining elements are not superconducting
is also explained by BCS-Eliashberg theory. However it should be kept in mind that many predictions of BCS theory
are not dependent on whether the pairing mechanism is the electron-phonon interaction or some other
boson exchange mechanism.  

In the previous section we discussed how the deviations from the BCS gap ratio $2\Delta/k_B T_c=3.53$
are explained within Eliashberg theory, and Figure~\ref{figure3} appeared to provide strong confirmation of the
validity of this interpretation. However, the theoretical steps to obtain both the horizontal and vertical coordinates
of each point in Fig.~\ref{figure3} are intertwined in a complicated way. It is interesting to redraw Fig.~\ref{figure3} using only
experimental data. In place of $\omega_{ln}$ we use the Debye temperature for the horizontal coordinate
and for the vertical coordinate we use the experimental values for the gap ratio, both quantities as given 
in Ref. \cite{poole2}. The results are shown in Fig. 6. It is not obvious  from Fig. 6 that
there is a simple relation between the gap ratio, the critical temperature and an average phonon
frequency represented here by the Debye temperature. The reason for the qualitatively different
behavior seen in Figs. 6 and \ref{figure3} is unclear \cite{remark2}.

%fig. 6
 \begin{figure}
%\resizebox{7.5cm}{!}{\includegraphics[width=5cm]{gaprat.pdf}}
\resizebox{7.5cm}{!}{\includegraphics[width=5cm]{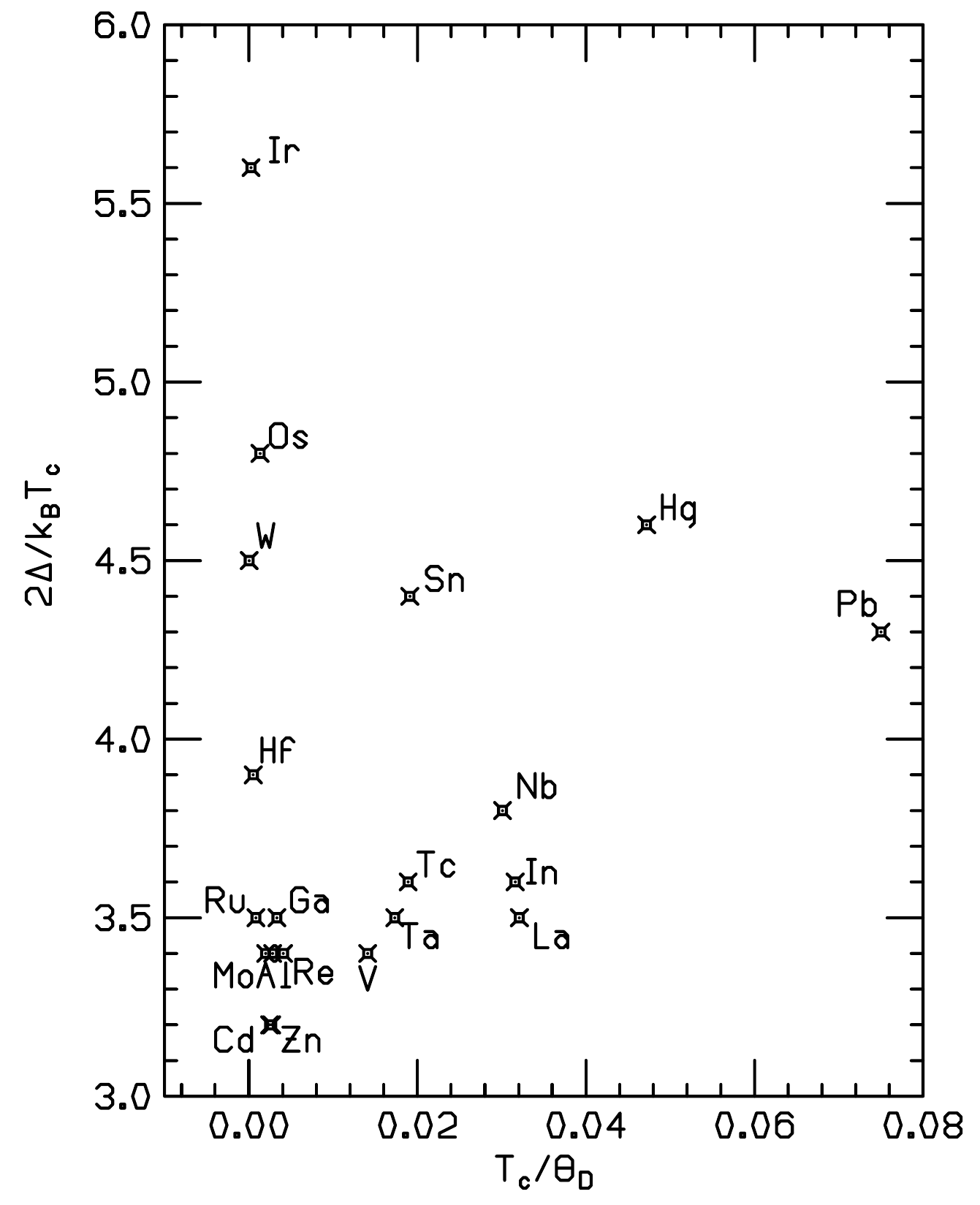}}
\caption { The gap ratio $2\Delta_0/(k_BT_c)$ as a function of $T_c/\theta _D$.  Note the considerable 
deviation  from the simple behavior shown in Fig.~\ref{figure3}. All data is taken from Ref. [\onlinecite{poole2}].}
\label{figure6}
\end{figure}

There is in principle a well-defined procedure to calculate the critical temperature of an element from first principles BCS-Eliashberg theory. given its lattice structure.
One needs to know the Fermi surface, the matrix elements of the electron-phonon interaction and the phonon dispersion curves, to find the parameters that go into the Eliashberg equation.
The electronic properties can be obtained from the modern theory of electronic structure of materials based on density functional theory. 
The phonon dispersion curves are usually obtained from a Born-Von Karman fit
to measured phonon frequencies, or alternatively from first principles.  However there are many subtleties involved in these calculations.
Examples of attempts to explain theoretically the observed critical temperatures of the elements are discussed in what follows.  
 
In an early contribution \cite{carbotte67}, Carbotte and Dynes computed the transition temperature of Al using as input inelastic neutron scattering data on phonons and the
Heine-Abarenkov pseudopotential for the electron-ion form factor. Solving the Eliashberg gap equation and assuming the weak coupling BCS relation
$2\Delta_0/(k_BT_c)=3.53$ they obtained a critical temperature $T_c=1.17$ K, in remarkable agreement with the experimental value $T_c=1.18$ K.
Using the same scheme the authors predicted \cite{carbotte68} that the critical temperature of Na and K should be much less than $10^{-5}$ K, and that the energy gap in
Pb is $\Delta_0=1.49$ meV, in good agreement with the measured value $1.35$ meV.

Using a similar first-principles approach, Allen and Cohen \cite{ac} computed the transition temperature of sixteen simple metals plus Ca, Sr and Ba. They used an 
isotropic model for the Fermi surface and the phonon spectrum, a Debye sphere for the phonon Brillouin zone, and a variety of different pseudopotentials.
They found that the calculated electron-phonon coupling $\lambda$ and resulting $T_c$ is quite sensitive to the details of the pseudopotential, and that
the results also depend on the assumed value of the band mass which is quite sensitive to the type of band calculation and form of the pseudopotential used. In addition
the results depend on the assumed value of $\mu^*$ which according to these authors may vary considerably from metal to metal and for which it is difficult to get
reliable first principles values.
The calculated values of the transition temperatures were found to be surprisingly good in view of all these uncertainties. The results for Pb, Sn, Tl, Hg and Zn were in reasonable
agreement with experiment (within a factor of 2). Large disagreement was found for the case of Ga, for which the calculations predicted $T_c<0.05$ K versus the experimental
value $T_c=1.09$ K. This was attributed to a failure of the spherical extended zone approximation used for the phonons \cite{ac}. However for the case of Sn the same effect was found
to give too large a value of $\lambda$ and $T_c$. For $Li$ and $Mg$ the critical temperatures were estimated to be around $1$ K and $10-80$ mK respectively. 
The paper concluded by urging that Mg and Li be tested for superconductivity, stating that ``The discovery of superconductivity in these materials would be a rather
convincing demonstration that the theory of the transition temperature had come of age.''

Motivated by this prediction an experimental attempt  to test for superconductivity in  $Li$ and $Mg$ down to $4$ mK was made shortly
thereafter \cite{limg}, with negative results. Several decades later superconductivity in Li at ambient pressure was detected at $0.4$ mK \cite{liexp}.
More sophisticated theoretical studies have not been able to resolve the discrepancy for Li \cite{cohen1,cohen2}, necessitating the assumption of a Coulomb pseudopotential as large as 
 $\mu^*=0.21$ \cite{cohen2}, much larger than the canonical value $\mu^*=0.1$, to account for the observed low $T_c$. 
 $Mg$ has not yet been found to be superconducting at any temperature. 
 
 In another study \cite{papa}, Papaconstantopoulos and coworkers calculated the critical temperature of the 32 metallic elements with $Z\leq 49$ using a theory of the
 electron-phonon interaction formulated by Gaspari and Gyorffy \cite{gas} for a rigid muffin-tin model, using experimental values for the Debye temperature obtained from specific
 heat measurements. $T_c$ was calculated from the McMillan formula using an empirical formula for the Coulomb pseudopotential that only depends on the density of states
 at the Fermi energy. To get better agreement with experiment, the contribution to the electron-phonon interaction arising from d-f scattering was reduced by a factor of 2
 from its first principles value. 
 
 The values found \cite{papa} for the critical temperature of Nb and V were 8.77 K and 4.62 K, in good agreement with the experimental values 9.2 K and 5.43 K. Also good agreement was
 found for Ti, $T_c=0.28$ K versus the experimental value $T_c^{exp}=0.39$ K and for Zr, $T_c=1.53$ K vs $T_c^{exp}=0.53$ K. However, many discrepancies were found:
 For technetium, $T_c=0.03$ K vs $T_c^{exp}=7.73$ K, for In, $T_c=0.04$ K vs $T_c^{exp}=3.40$ K, for Ru, $T_c=0$ vs $T_c^{exp}=0.49$ K, for Mo, $T_c=0$ vs $T_c^{exp}=0.92$ K,
 for Ga, $T_c=0$ vs $T_c^{exp}=1.08$ K, for Zn, $T_c=0$ vs $T_c^{exp}=0.375$ K, for Sc, $T_c=0.51$ K vs $T_c^{exp}=0$, for Al, $T_c=0$ vs $T_c^{exp}=1.18$ K,
  and for Li, $T_c=0.65$ K vs $T_c^{exp}=0.0004$ K. Nevertheless the authors concluded that their method can reliably account for all the high temperature superconductors in the first
  half of the periodic table, and viewed this as a promising step in the direction of predicting new superconductors in more complex materials \cite{papa}.
  
  In a similar calculation for Pb \cite{papa2}, the authors found an ab-initio value for $\lambda$ which was half the value found experimentally from tunneling
  experiments. They  argued that for Pb the rigid muffin-tin model has to be corrected and proposed a correction term to the rigid muffin tin potential. Imposing the constraint that
  its Fourier transform  of this term yields the correct limit as the wavevector $q\rightarrow 0$ they obtained a renormalized $\lambda$ which was in excellent agreement with experiment.
  
  An ab initio calculation of superconducting transition temperatures  using the rigid muffin tin approximation 
  was performed by  Glotzel,  Rainer and  Schobern   \cite{abinitio}, using for the lattice dynamics a Born-von Karman model fitted to measured
  phonon frequencies, for the elements $V$, $Nb$, $Ta$, $Mo$, $W$, $Pd$, $Pt$, $Pb$. The calculated versus experimental (in parentheses) values of $T_c$, in K, were
  21.4 (5.4), 17.4 (9.2), 9.2 (4.4), 0.8 (0.91), 0.07 (0.015), 1.4 (0), 3.2 (0), 2.6  (7.2).
   The authors concluded that at the present state of the art (year 1979) ab initio theory was incapable of producing reliable values of $T_c$.

  The papers discussed above \cite{carbotte68,carbotte67,ac,papa,papa2,abinitio} are among the most prominent early attempts to calculate $T_c$'s of elements from first 
  principles. To learn what has been achieved since then in that respect  we looked at all  the papers citing these seminal works. 
  There are a few more recent calculations of $T_c$'s of elements that report improved agreement with 
  experiment \cite{tc1,tc2,tc3,tc4,tc5,tc6,tc7,tc8,tc9}. However,  by and large the interest of the leading practitioners of this science/art and their disciples
  shifted to calculate critical temperatures of more complicated materials, some of which will be discussed in other papers in this Special Issue.
  As a consequence, we face the somewhat disconcerting situation that the calculation of critical temperatures of   the simplest materials, the elements at ambient pressure,
  within conventional BCS-Eliashberg theory, does not seem to be developed to a stage where it can predict the observed $T_c$ from first principles.
  This situation, recognized and termed ``superflexibility'' by D. Rainer back in 1982 \cite{rainer}, does not appear to have been resolved since then,
  despite recent claims to the contrary \cite{gross}.

  \section{superconductivity in alloys and  simple compounds}
Essentially all elements, whether superconducting or not, make superconducting 
alloys and compounds when combined with one or two other elements. The large majority of these superconductors
are believed to be conventional superconductors. 

A large number of superconducting alloys have been investigated, as surveyed by Matthias, Geballe and
Compton \cite{mgt}. Alloys can have $T_c$'s that are higher or lower than those of its constituents. For example, addition of $20-30\%$ $Zr$ ($T_c=1.1K$) to $Nb$ ($T_c=9.2K$) raises its critical temperature to $11K$, while $8\%$ of $Sn$ dissolved into $Nb$ lowers its
$T_c$ to $5.6K$. It 
is often the case that the  $T_c$ of an alloy bears little relation with that of its constituting elements,
for example, $30\%$      $W$ ($T_c=12mK$) dissolved in $Pt$ (non-superconductor) is superconducting with $T_c=0.40K$,
$25\%$ of $Re$ ($T_c=1.4K$) in $W$ raises its $T_c$ to $4.2K$, etc.
Thousands of intermetallic compounds as well as carbides, nitrides, oxides, sulfides, hydrides, etc, in a large variety of different
crystal structures have been studied and many found to be superconducting. References \cite{roberts,poole,vonsovsky} survey many of these materials.

One such simple class is that consisting of binary compounds with a metallic and a
non-metallic atom forming a sodium-chloride structure.
Another simple class are binary intermetallic compounds with a cesium-chloride structure. Other
examples are Laves phases, 
metallic $AB_2$ type compounds in cubic or hexagonal structures, several of which are superconducting.  Examples of these compounds, as well as of technologically important substitutional alloys with the bcc structure,
with their $T_c$'s and
 values of the upper critical field, are shown in Table IV.

  There have   been several  calculations and predictions of critical temperatures of such simple compounds based on the BCS-Eliashberg formalism,
  with mixed success. For example, for $VN$, $NbN$ and $TaN$, first principles calculations yielded \cite{vnetc} $Tc$'s $19.7K$, $17.1K$ and $14.6K$, in reasonable  agreement with the
  experimental values $9,25K$, $17K$ and $8.9K$. However, using the same methodology it was
  predicted \cite{mon} that $MoN$ if it formed in the sodium-chloride structure would have a surprisingly high $T_c\sim 29K$.    When experimentalists succeeded in stabilizing this structure in $MoN$ films, the superconducting transition temperature was found to be only around $3K$ \cite{monexp}.
  It was proposed that the discrepancy might be due to the presence of substantial disorder in the films \cite{vnetc}.
More recent calculations for $NbC$, $NbN$ and $NbC_{1-x}N_x$ alloys  \cite{nbnrec} found that
  Fermi surface nesting and the associated Kohn anomaly greatly increases the electron-phonon coupling thus accounting  for the relatively high $T_c$ of these materials. 
    
    For the carbides $NbC$, $TaC$, and $HfC$ first principles calculations yielded \cite{carbtheo} $T_c$ values
  $10.8K$, $9.6K$ and $0$, in good agreement with the experimental values $11.1K$, $11.4K$ and $0$. A more recent calculation
   for a variety of carbides   found that Fermi surface nesting plays a significant role in
  enhancing $T_c$ \cite{carb}.

  For the cubic Laves phase compounds $ZrV_2$, $ZrCo_2$, and $ZrFe_2$, first principles calculations of the superconducting transition temperatures \cite{lavestheo} yielded  
  the values $17K$, $0K$ and $9K$, for experimental values $9K$, $0K$ and $0K$. The discrepancy for $ZrFe_2$ is explained by the fact that the material is a ferromagnet
  while in the calculation a paramagnetic state is assumed. 
  
  \begin{table}
\caption{Some compounds and alloys with simple structures and their critical temperatures,
and some $H_{c2}$ values with $T_{mess}$ the temperature at which
$H_{c2}$ was measured ((0) means extrapolated to zero temperature). See, for example, Ref. \onlinecite{roberts}.}
\begin{tabular}{l | c | c | c | c  }
  \hline
    \hline
Structure  & Material & $T_c (K)  $   & $H{c2}(kOe) $     \cr
 \hline  \hline
Cubic NaCl & MoC & 14.3& 52 (4.2)   \cr \hline
\: \: \: \: " & VN & 9.25& $>$250 (4.2)   \cr \hline
\: \: \: \: " & NbN & 17& $>$250 (4.2)   \cr \hline
\: \: \: \: " & TaN & 8.9& $>$250 (4.2)   \cr \hline
\: \: \: \: " & NbC & 11.1& 16.9 (4.2)   \cr \hline
\: \: \: \: " & NbO &1.4&     \cr\hline
\: \: \: \: " & ZrB&3.4&     \cr \hline
\: \: \: \: " & ThS &0.5&     \cr \hline
\: \: \: \: " & ThSe &1.7&     \cr \hline 
\: \: \: \: " & TaC &11.4&   4.6 (1.2)  \cr \hline
\: \: \: \: " & TeGe &0.4&     \cr \hline
\: \: \: \: " & LaS &0.9&     \cr \hline
\: \: \: \: " & PdH &9.6&     \cr \hline \hline
Cubic CsCl & CuSc &0.5&     \cr \hline
\: \: \: \: " & CuY &0.3&     \cr \hline
\: \: \: \: " & AgY&0.3&     \cr \hline
\: \: \: \: " & AgLa&0.9&     \cr \hline
\: \: \: \: " & AgSc &2&     \cr \hline \hline
Laves cubic & CaRh$_2$ &6.4&     \cr 
or hexagonal&  & &     \cr \hline
\: \: \: \: " & CaIr$_2$&2&     \cr \hline
\: \: \: \: " & ScRu$_2$&2&     \cr \hline
\: \: \: \: " & ScOs$_2$&2&     \cr \hline
\: \: \: \: " & ZrV$_2$&9&  103 (4.2)    \cr \hline
\: \: \: \: " & HfV$_2$&2&   200 (4.2)  \cr \hline
\: \: \: \: " & AgY &0.3&     \cr \hline \hline
Bcc alloys& Mo$_x$Re$_{1-x}$&11.8&  27.9 (1.3)   \cr \hline
\: \: \: \: "& Nb$_x$Ta$_{1-x}$&9&  8.7 (0)   \cr \hline
\: \: \: \: "& Nb$_x$Ti$_{1-x}$&9.9&    141 (0) \cr \hline
\: \: \: \: "& Nb$_x$Zr$_{1-x}$&11.1&  103 (0)   \cr \hline  
 \hline
 \end{tabular}
\end{table}

\section{beyond bcs theory}
While the BCS-Eliashberg formalism can often account for observed critical temperatures through detailed calculations as reviewed above, it does not provide simple criteria to understand why critical temperatures
are sometimes high, sometimes low, and sometimes zero, neither  for the elements, alloys
 and simple compounds discussed here nor for other classes of materials discussed in this Special Issue. For example, this state of affairs is
acknowledged in a recent study of superconductivity of elements under high pressure \cite{pressure}, where the authors state that even though
``it has become clear that strong electron-phonon coupling can account for the remarkable superconductivity of Y under pressure'',
``What is lacking is even a rudimentary physical picture for what distinguishes Y and Li ($T_c$ around 20K under pressure) from other elemental metals
which show low, or vanishingly small, values of $T_c$''. We suggest that the same statement applies to the elements,
alloys and simple compounds at ambient pressure discussed in this article. 
For this reason it is of interest to mention briefly some empirical criteria that have been used 
to understand the presence or absence of superconductivity and/or the magnitude of critical temperatures
in elements and simple compounds that do not rely on BCS-Eliashberg theory.

As discussed elsewhere in this Special Issue \cite{geballe}, B. Matthias proposed certain rules (``Matthias' rules'') to understand the behavior of $T_c$ in alloys of transition metals  \cite{mr}, pointing out 
that the critical temperature appears to depend solely on the average number of electrons per atom (e/a ratio). 
An explanation of this e/a dependence based on conventional BCS theory is given in Ref. \cite{pines}, and an alternative explanation is proposed
in Ref. \cite{tm}. 
Matthias also noted that simple cubic and hexagonal structures are favorable for superconductivity \cite{matthias57}.
See ref. \cite{geballe} for further discussion. Another Matthias' insight,  that may \cite{mcmillan68,testardi} or may not \cite{coulomb} be related to
BCS theory, was that \cite{instab}  ``Crystallographic instabilities seem to be a necessary condition for high superconducting transition
temperatures in multicomponent phases''.

As mentioned in the introduction, among the earliest superconducting compounds investigated were $CuS$ and $PbS$ (see table I). In 1932, 
Kikoin and Lasarew pointed out \cite{kikoin} that the Hall coefficient of these materials was particularly small, compared to that
of other similar semiconductors that were not superconductors. They wondered whether the small value of the Hall coefficient was related to
the existence of superconductivity.
Tabulating the values of $R$ (Hall coefficient) and $R\sigma$ ($\sigma=$ electrical conductivity) for several superconducting elements and some
binary compounds known at the time, they found that superconductivity was strongly correlated with small values of $R$ and
particularly with small values  of $R\sigma$.  

Later, Linde and Rapp pointed out \cite{linde} that for many non-transition metal alloys the  critical temperature increases as the  Hall coefficient 
decreases as a function of composition, at the same time as the electron-phonon coupling as inferred from the temperature derivative of the resistivity is
increasing. Examples of these systems are $AuGa$, $AuAl$, $AuGe$, $AuZn$, $AuSn$ and $AuIn$. In 25 out of 27 alloy systems considered they found this correlation.  

In a series of papers, Chapnik pointed out \cite{chapnik1,chapnik2,chapnik3,chapnik4} that in fact superconductivity is correlated with a positive sign of the Hall coefficient in a large number of elements, alloys and compounds. For example, he pointed out that $Au$ and $Pd-Ag$ alloys with a cubic crystal structure (usually favorable to superconductivity) and a negative Hall coefficient
are not superconducting \cite{schuller}.
Chapnik explained
 the observation of Linde and Rapp with a two-band model where the decrease of $R$ pointed out by Linde and Rapp would result
from an increasing hole concentration. 

One of the present authors examined correlations between 13 normal state properties of elements and superconductivity \cite{correl} from a statistical point of view. It was
 found that properties assumed to be important within BCS theory rank low in predictive power
regarding whether a material is or is not a superconductor. Instead, properties with highest predictive power in
this respect were found to be bulk modulus, work function and particularly Hall coefficient as pointed out by Chapnik. These properties play no special role within BCS theory. The correlation  of $T_c$ with Hall coefficient for the elements is shown in Fig. 7.

 \begin{figure}
%\resizebox{7.5cm}{!}{\includegraphics[width=5cm]{gaprat.pdf}}
\resizebox{8.5cm}{!}{\includegraphics[width=5cm]{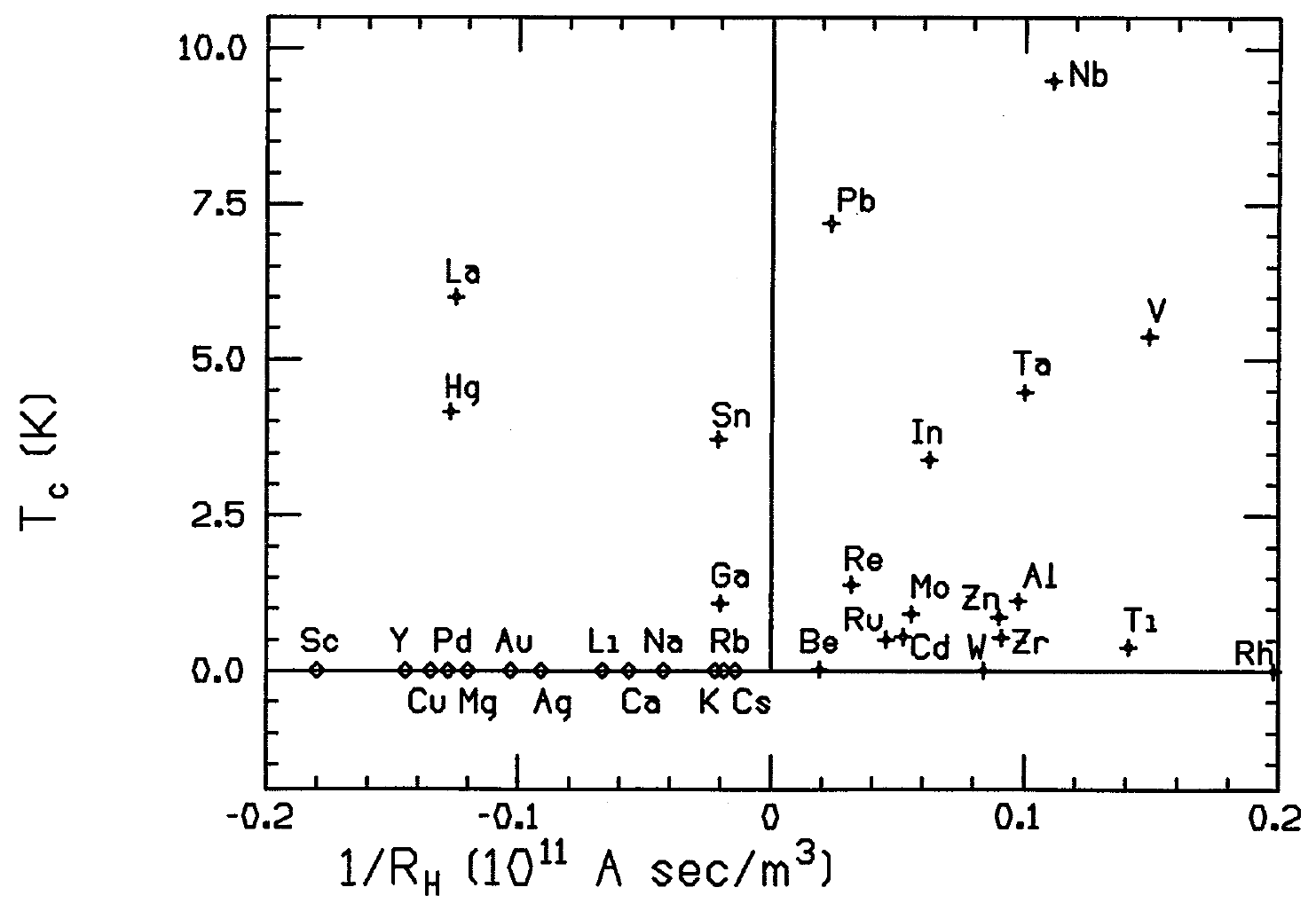}}
\caption { Superconducting critical temperature of the elements
plotted versus the inverse Hall coefficient at low temperatures and high fields. Note that superconductivity is predominantly associated with a 
$positive$ Hall coefficient.}
\label{figure7}
\end{figure}

Another early empirical observation was made by Meissner and Schubert \cite{ms,justi}. They pointed out that the volume per valence electron (the difference between the atomic and ionic
volume, divided by the number of conduction electrons per atom) is particularly small in superconducting elements compared to non-superconducting elements, with the smallest
values associated with the highest transition temperatures. It is interesting that this criterion gives a qualitative understanding for why high critical temperatures are often
achieved under high pressures, as discussed in several other papers in this Special Issue.

\section{summary and discussion}

%In this article we gave a brief review of superconductivity in elements, alloys and simple compounds at ambient pressure. These materials are generally believed to be described by 
%the conventional BCS-Eliashberg theory, with the superconductivity caused by an effective electron-electron attraction resulting from the electron-phonon interaction, that overcomes the
%repulsive Coulomb interaction between electrons. The resulting superconducting state is s-wave, and the magnitude of the critical temperature is limited by the fact that phonon energy scales are much lower than electronic energy scales. 
%The same theoretical framework is generally believed to explain why many elements, alloys and simple compounds do
%$not$ become superconducting at any temperature.

In this article we gave a brief review of superconductivity in elements, alloys and simple compounds at ambient pressure. These materials are generally believed to be described by 
the conventional BCS-Eliashberg theory, with the superconductivity caused by an effective electron-electron attraction resulting from the electron-phonon interaction, that overcomes the
repulsive Coulomb interaction between electrons. The resulting superconducting state is s-wave, and the magnitude of the critical temperature is limited by the fact that phonon energy scales are much lower than electronic energy scales. 
The same theoretical framework is generally believed to explain why many elements, alloys and simple compounds do
$not$ become superconducting at any temperature.

However, this raises the question: why are none of the non-conventional mechanisms proposed to apply to other
classes of materials discussed in this Special Issue  operative in the class of superconductors discussed in this article?

For example, it has been argued that spin fluctuations induced by strong Coulomb repulsion
prevent conventional superconductivity from occurring in  
$Sc$ and $Pd$ \cite{pd}. Why isn't a spin-fluctuation mechanism \cite{scalreview} proposed to be operative in several of the other classes of materials
discussed in this Special Issue  such as cuprates, pnictides, heavy fermions, Pu compounds, layered nitrides, 
organics, cobaltates and $Sr_2RuO_4$, 
 operative in $Sc$ and $Pd$  and gives rise to superconductivity in them or in alloys or simple binary compounds
 with $Sc$ or $Pd$ as one of the components? Or, why doesn't the $s\pm$ mechanism proposed to operate in iron pnictides
operate in simple compounds that also have both hole-like and electron-like pieces to the Fermi surface?

We suggest that the question why none of the elements, alloys and simple compounds can take advantage
of any of the non-conventional mechanisms operating in other materials is worth pondering, and
that finding its answer could significantly advance our understanding of superconductivity in materials.

We also suggest that given the significant advances that have taken place in recent years in first principles
calculations of electronic properties of materials \cite{gross,new1,new2},  it should be possible using 
BCS-Eliashberg theory
to better account for the $T_c$'s measured in elements, alloys and simple compounds, as well as the non-existence of superconductivity
in many of these materials, than what was recounted in Sects. III and IV.
 For example, the theory is claimed to reproduce the $T_c=39K$ of $MgB_2$ from first
principles to within $10\%$ without adjustable parameters \cite{liu,choi,floris, margine},
in rather complicated calculations where anharmonicity and anisotropy of the phonon spectrum is fully taken into account.
It should be simpler and at least as successful to apply these techniques to elements and simple compounds.
For a handful of elements and   simple compounds this has recently  been done and claimed to successfully reproduce the measured $T_c$'s \cite{gross,margine,gross2,gross3}. 
It should be systematically done for many elements and simple compounds. For example, can these methods reproduce  the non-existence of superconductivity in the early
and late transition metal series (e.g. $Sc$, $Y$, $Pd$, $Pt$)  and the extremely low $T_c$ of $Li$  without additional ad-hoc assumptions such as a 
large $\mu^*$ as was done in the past \cite{pd,papa,cohen2,ashcr}? Can one computer program designed to calculate $T_c$ of
binary compounds forming a cubic $NaCl$ structure such as the ones listed in Table IV, compute the critical
temperature (including $T_c=0$) of  binary compounds in such a structure by simply entering $Z_1$, $Z_2$ and $a$, the atomic
number of each constituent and the lattice constant, with $no$ further adjustments? Approximate agreement with experiment
for dozens of such elements and compounds  would be an impressive validation of 
BCS-Eliashberg theory as the correct theory for the description of the superconductivity of
conventional superconductors. On the other hand, significant disagreement would suggest that something is
amiss with the present understanding of the validity of  BCS-Eliashberg theory to describe superconductivity
in simple materials \cite{validity}.


\begin{references}
\bibitem{remark0} A third critical field, $H_{c3} > H_{c2}$, is also possible, predicted \cite{saint-james63} to occur
at the surface of a type-II superconductor. This was verified experimentally for several type-II materials \cite{webb68,karasik70}
and even for a type-I elemental superconductor \cite{mcevoy67}.
\bibitem{remark1} Note that retardation effects are taken into account in BCS theory in a very phenomenological
way through the imposed cutoff. This cutoff is imposed in momentum space (not in frequency space), and
because of the Fermi Liquid nature of BCS theory \cite{rainer86} this can be rewritten as a formulation in
terms of Matsubara frequencies with a cutoff in this space, so it better resembles Eliashberg theory.
\bibitem{remark2} Also note that there are discrepancies in the reported experimental values of, for example, $2\Delta_0/(k_BT_c)$, in the
literature.
\end{references}
\end{document}